\documentclass[global,twocolumn]{svjour}
\usepackage{epsfig}
\usepackage{graphicx}
\usepackage{amsmath}
\usepackage{amsfonts}
\usepackage{amssymb}
\journalname{Applied Physics B}

\begin{document}

\title{A broadband Ytterbium-doped tunable fiber laser for $^{3}\mathrm{He}$ optical
pumping at 1083~nm}
\author{G. Tastevin\inst{1}\thanks{electronic mail : tastevin@lkb.ens.fr, Fax : 33-1-44323434}
\and S. \ Grot\inst{2}
\and E. Courtade\inst{1,3}
\and S.\ Bordais\inst{2}
\and P.-J. Nacher\inst{1}}

\institute{Laboratoire Kastler Brossel\thanks
{Research laboratory affiliated to the Universit\'e Pierre et Marie Curie and to the Ecole Normale Sup\'erieure,
associated to the Centre National de la Recherche Scientifique (UMR~8552)},
24 rue Lhomond, 75231 Paris cedex 05, France.
\and
Keospsys S.A., Department of Research and Development, 21 rue Louis de Broglie,
22300 Lannion, France.
\and
Present address : Dipartimento di Fisica, Universit\`a di Pisa, Via Buonarroti 2, I-56127 Pisa Italy.
}

\date{\today}
\maketitle
\begin{abstract}
Large amounts of hyperpolarized $^3$He gas with high nuclear polarization rates are required 
for use in neutron spin filters or nuclear magnetic resonance imaging of human lung. 
Very high efficiency can be obtained by metastability exchange optical pumping using 
multimode lasers to excite the 2$^3$S-2$^3$P transition at 1083~nm. Broadband 
Ytterbium-doped tunable fiber lasers have been designed for that particular application. 
Different options for the architecture of the fiber oscillator are presented and compared. Emphasis
is given to a linear cavity configuration that includes a high reflectivity fiber mirror 
and a low reflectivity tunable fiber Bragg grating. Optical 
measurements are performed to finely characterize the spectral behavior of the lasers. 
Atomic response is also quantitatively probed to assess the optimal design of the 
oscillator for optical pumping. Multimode operation matching the 2~GHz Doppler-broadened 
helium resonance line and tunability over more than 200~GHz are demonstrated. Boosting the 
output of this fiber laser with a Yb-doped fiber power 
amplifier, all-fiber devices are built to provide robust, high power turn-key 
sources at 1083~nm for improved production of laser polarized $^3$He.

\vspace{5mm}\hskip-\parindent{\bfseries PACS}\hskip\parindent42.60.-v;
42.55.Wd; 32.80.Bx
\end{abstract}

%Received: date / Revised version: date}
%

\titlerunning{A broadband Yb-doped tunable fiber laser for $^{3}\mathrm{He}$ optical pumping at 1083~nm}

\section{Introduction\label{section1}}

Laser-polarized $^{3}\mathrm{He}$ is used in a variety of research fields,
ranging from low-temperature to nuclear physics \cite{Leduc90,Becker98,Rohe99}%
. It has emerged in the past decade as a very promising tool for applications
like the preparation of spin filters for cold neutrons \cite{Becker98,Jones00}%
, or magnetic resonance imaging of air spaces in human lungs
\cite{Tastevin00,Chupp01,Kauczor01}. This has strongly increased the demand
for production of large amounts of gas with very high nuclear polarizations
and high production rates. The most efficient method to meet both requirements
is metastability exchange optical pumping (MEOP). It relies on optical pumping
(OP) of the $2^{3}\mathrm{S}$\ metastable state of helium with 1083~nm
resonant light \cite{Colegrove63,Nacher85}. Large nuclear polarizations can be
prepared in a low-pressure gas at room temperature within a few seconds
\cite{Nacher85,Leduc00}. Polarization-preserving compression is performed
after OP to obtain dense gas, with a choice of possible schemes depending on
the targeted application \cite{Becker94,Nacher99,Gentile00}. Home-built
1083~nm light sources \cite{Leduc90,Aminoff89} are now replaced by commercial
lasers \cite{Leduc00,Stoltz96,Chernikov97,Mueller01}.\ Their development may
lead to improved performances and open the way to a wider dissemination of
this laser-based polarization technique.

For massive production of highly polarized $^{3}\mathrm{He}$ gas powerful
amplifiers have been built, a progress due to the rapid development of
fiber-based technology for telecommunications. They are operated in a
conventional master oscillator power fiber amplifier configuration (MOPFA),
with a seed laser providing the desired 1083~nm radiation (a DBR laser diode
\cite{Chernikov97,Mueller01} or a fiber laser \cite{Leduc00}). OP rates are
crucially determined by the number of atomic transitions induced per unit
time, which depends both on the available power intensity and on the spectral
characteristics of the laser emission. The laser cavity must hence be
carefully optimized to efficiently match the spectral power distribution
profile to the Doppler broadened atomic line.

The main objective of this article is to describe all-fiber 1083~nm laser
oscillators specially designed for helium OP. To finely characterize their
spectral characteristics and assess their performances a series of tests and
measurements is performed. Absolute OP performances are actually difficult to
determine because experimental results strongly depend on the choice of
operating conditions. The second objective of this article is to present in
detail the basic tests performed in order to provide reference tools that can
be used to quantitatively characterize and appropriately compare all 1083~nm
sources available for helium OP. Section~\ref{section2} briefly compiles the
relevant features of the $2^{3}\mathrm{S}$-$2^{3}\mathrm{P}$ transition and
describes the currently used OP schemes to specify the main laser
requirements. Section~\ref{section3} describes the architecture of the fiber
oscillators and the performed optical measurements. Section~\ref{section4}
focuses on the measurements of the atomic response to the laser spectral
characteristics that provide a guide to optimal design of fiber lasers
tailored to OP applications.

\section{Metastability exchange optical pumping and laser requirements}

\label{section2}MEOP is a polarization technique involving only helium atoms,
based on an indirect\ spin orientation process \cite{Colegrove63,Nacher85}.
Due to hyperfine coupling in the $2^{3}\mathrm{S}$ state, OP\ on the
electronic $2^{3}\mathrm{S}$-$2^{3}\mathrm{P}$ transition induces partial
orientation of the nuclei of the metastable atoms. Metastability exchange
occurs during collisions in which an incoming metastable atom transfers its
electronic excitation energy to an incoming ground state atom, leaving the
outgoing ground state atom with the partially polarized nucleus.

\begin{figure}[th]
%h=here, t=top, b=bottom, p=separate figure page
%
%
%
%
%
%
%
%
%
%
%
%
%
%
%
%
%
%
%
%
%
%
%
%
%
%
%
%
%
%
%
%
%
%
%
%
%
%
%
%
%
%
%
%
%
%
%
%
%
\par
\begin{center}
\leavevmode
\includegraphics[ keepaspectratio,width=0.95\linewidth,clip= ]{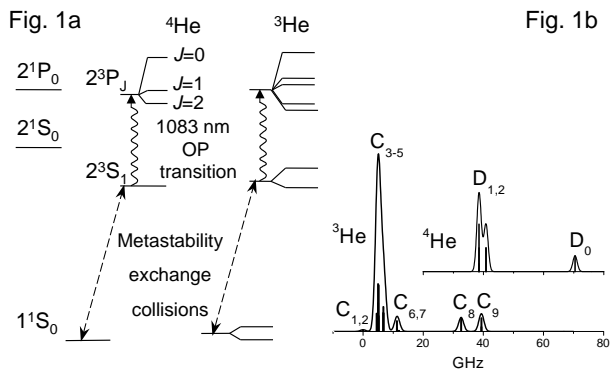}
\end{center}
\caption{a: Low energy states of the $\mathrm{He}$ isotopes used for MEOP. b:
Isotopic shift, fine and hyperfine structure of the $2^{3}\mathrm{S}$%
-$2^{3}\mathrm{P}$ transition lead to a 70~GHz frequency spread at null
magnetic field.}%
\label{figure1}%
\end{figure}

The structure of the low lying energy states of helium and of the
$2^{3}\mathrm{S}$-$2^{3}\mathrm{P}$ transition at null magnetic field is
indicated in Fig.~\ref{figure1}. More details and numbers can be found in \cite{Courtade02} and
references therein. The atomic resonance lines are distributed over nearly 70
GHz, with line widths originating from various processes. The radiative decay
rate of the $2^{3}\mathrm{P}$ state is 1.02$\times$10$^{7}$~$\mathrm{s}^{-1}$.
Atomic collisions introduce a pressure-dependent contribution (10$^{8}%
$~$\mathrm{s}^{-1}$/mbar). For gas confined in an OP cell, the atomic lines
are strongly Doppler broadened at room temperature. The Doppler full width
half maximum\ (FWHM) $D$ is on the order of 2 GHz for the helium atom (at
300~K $D$=1.98~GHz FWHM for $^{3}\mathrm{He}$, 1.72~GHz FWHM for
$^{4}\mathrm{He}$).

MEOP is usually performed at low helium gas pressure (around 1~mbar) with a
low applied magnetic field (up to a few mT). For OP of pure $^{3}\mathrm{He}$
one advantageously uses one of the two resolved components, C$_{8}$ or C$_{9}$
(respectively the $2^{3}\mathrm{S,}$ $F$=1/2 - $2^{3}\mathrm{P}_{0}$ and
$2^{3}\mathrm{S,}$ $F$=3/2 - $2^{3}\mathrm{P}_{0}$ transitions)
\cite{Nacher85}. Isotopic mixtures may be efficiently pumped using the D$_{0}$
line ($2^{3}\mathrm{S}$ - $2^{3}\mathrm{P}_{0}$ transition of $^{4}%
\mathrm{He}$) \cite{Stoltz96}, significantly shifted from the $^{3}%
\mathrm{He}$ lines. To reach these three resonance lines the infrared pumping
laser must be tunable over 40~GHz. The low applied magnetic field is required
only to prevent fast magnetic relaxation and it has a negligible effect on the
structure of the atomic states. In particular all Zeeman energy splittings are
much smaller than the Doppler width and the pumping light must be circularly
polarized to selectively depopulate adequate sublevels and deposit angular
momentum into the gas. As nuclear polarization builds up the light absorption
strongly decreases due to the depletion of the population in the pumped atomic
sublevels. Simultaneously the absorption probability increases for light with
the opposite circular polarization, so that any residual fraction of the
pumping beam with that wrong circular polarization will very efficiently
contribute to reduce the ultimate nuclear orientation \cite{Nacher85,Leduc00}.
In practice the beam polarization (and hence the optical quality of all
elements along the light path including the OP cell windows) becomes
especially crucial at high laser intensities. A set of high-quality polarizing
cube and low order quaterwave plate is usually inserted in front of the
OP\ cell to get pure circular polarization. Therefore the laser light must
have a linear polarization whose long term stability directly determines that
of the effective pumping power.

Operation at high magnetic field has recently been shown to yield improved OP
performances at higher gas pressure \cite{Courtade00,Nacher02}. The structure
of the energy levels is strongly modified, and the various Zeeman sublevels
are no longer degenerate. This advantageously removes the stringent constraint
on the degree of\ light polarization for the pumping beam. However, it also
substantially extends the required tunability range for the laser source, to
nearly 150~GHz at 1.5~T, for instance \cite{Courtade02}.

Extensive studies of the MEOP process have been performed in sealed glass
cells over the past 40 years. In order to populate the $2^{3}\mathrm{S}%
$\ metastable state, a plasma discharge is sustained in the helium gas. In
short, the steady state density of $2^{3}\mathrm{S}$\ atoms (hence the OP
light absorption and angular momentum deposition rates in the gas) increases
with the plasma intensity. Yet various operating conditions (gas purity,
pressure, and cell dimensions) may set limits to the actual $2^{3}\mathrm{S}$
lifetime and density. The discharge also introduces substantial nuclear
relaxation due to collisions in the highly excited states, emission of
circularly polarized fluorescence light by the plasma \cite{Pinard74}, or
formation of metastable He$_{2}$ molecules at high pressures (a process
enhanced at high laser powers due to massive promotion of He atoms to the
$2^{3}\mathrm{P}$ state, where the cross section to produce these He$_{2}$
molecules is 100 times larger than in the $2^{3}\mathrm{S}$ state
\cite{Emmert88}). The most favorable plasma conditions leading to highest
polarizations but slow pumping rates, are usually found for weak discharges in
a very pure He gas at low pressure for transverse cell dimensions on the order
of a few centimeters. Indeed, the actual optimal plasma and pressure also
depend on OP cell shape and size (e.g., due to radiation trapping), as well as
on OP laser power and spectral characteristics \cite{Leduc00}.

The optimal operating conditions have recently been revisited due to the
development of applications where large quantities of highly polarized
$^{3}\mathrm{He}$ gas are needed. Most current schemes are based on gas flow
through one or several OP cells, where it gets optically pumped at low
pressure and mT magnetic field, and accumulation of compressed polarized gas
in appropriate storage vessels \cite{Becker94,Nacher99,Gentile01}. Cell
dimensions, materials, and gas flow rates are optimized to achieve the best
efficiency with the available laser source, which usually results from a
tradeoff between high nuclear polarizations and fast production rates. In
addition, actual OP conditions may\ not be ideal (gas purity, for instance, is
never as high as in carefully prepared sealed cells). Taking into account the
finite transit time of the atoms in the OP cell, pressure, atomic transition
line and discharge intensity are also often selected to shorten the OP time.
In all cases, the latter decreases at high photon fluxes, leading to improved
performance when powerful lasers are available. The need for very high laser
output powers is further increased by the rapid development of polarized
$^{3}\mathrm{He}$ applications, which requires either upscaling the existing
gas polarizers or significantly improving the overall production rates in
order to meet the growing demand for polarized gas.

MEOP is a very efficient and fast process. It typically yields 2 polarized
nuclei per absorbed photon at 1083~nm, but this efficiency is reduced at high
nuclear polarizations \cite{Leduc00}. Therefore, a strong increase in laser
power is required to slightly improve the steady state nuclear polarization
close to the limit set by the intrinsic non linear effects in OP
\cite{Nacher85}.

To remain efficient at high laser powers and minimize line saturation effects,
OP must involve the largest possible number of $2^{3}\mathrm{S}$ atoms, i.e.,
optimally interact simultaneously with all velocity classes. A singlemode
laser is absorbed by atoms from a single velocity class, whose
pressure-broadened line width $\delta$ determines the fraction of
$2^{3}\mathrm{S}$ atoms actually in resonance with the laser: $X_{\mathrm{s}}%
$=$\delta$/$D$ (typically 1~\% at 1~mbar for pure $^{3}\mathrm{He}$ gas at
room temperature). The per-atom photon absorption probability (proportional to
the power density, i.e., the light intensity per unit area) is equal to the
spontaneous emission rate for a critical laser power density $\mathcal{P}%
_{crit.}=2\omega^{3}\hbar DX_{\mathrm{s}}/3\sqrt{\pi}c^{2}T_{ij}$
=~$0.277X_{\mathrm{s}}/T_{ij}$~(W/cm$^{2}$), where $\omega$ is 2$\pi$ times
the resonance frequency and $\hbar$ $1/2\pi$ times the Planck constant, $c$
the velocity of light, and $T_{ij}$ the transition matrix element between the
relevant atomic sublevels $i$ and $j$ \cite{Nacher85}. It is on the order of
10~mW/cm$^{2}$ for a circularly polarized singlemode laser ($X_{\mathrm{s}}%
$=1~\%) tuned to the C$_{8}$ line ($T_{ij}$~= 0.29) and 1~W/cm$^{2}$ for a
broadband laser emitting over the entire Doppler line width ($X_{\mathrm{s}}%
$=1). A more quantitative description of the impact of optical saturation on
the OP dynamics can be derived from a realistic and detailed model of the OP
process and from computer calculations taking into account the spectral and
geometric features of the laser beam \cite{Leduc00}. A single phenomenological
parameter ($X_{\mathrm{s}}$) is used to specify the spectral coverage between
the laser and Doppler profiles. This parameter is mainly determined by the
laser line width and mode structure. Collisional redistribution between
velocity classes also plays a role \cite{Courtade02}, but it is only crudely
taken into account for the numerical simulations \cite{Leduc00}. Attempts to
accurately reproduce saturation profiles measured for narrowband laser sources
bring to light the need to improve such input parameters, and maybe also to
include other effects like radiative broadening of resonance lines (at high
laser powers) \cite{CourtadeTh}. For a highly multimode laser, an optimal
spectral coverage is expected to exist: on the one hand the number of atoms
significantly excited by resonant light within the Doppler profile increases
with the laser line width; on the other hand the laser efficiency decreases
for an excessively broad emission spectrum, due to reduced absorption rates in
the wings of the Doppler atomic line. Assuming for instance a continuous
Gaussian laser frequency distribution of width $L$, the fraction of atoms
submitted to more than half of the incident power varies as $erf(\sqrt
{Ln(2)}L/D)$; it starts growing almost linearly with $L/D$ then rises more
slowly, exceeding 90\% for $L/D>1.4$. In contrast the ratio of the total
absorbed laser power to the incident one scales like $1/\sqrt{1+L^{2}/D^{2}}$;
it becomes for instance less than 50\% for $L/D>$ $1.7$. Therefore, the
optimal coverage may be qualitatively inferred to lie around $L\simeq D$.

In summary, the OP rates basically depend on the number of atomic transitions
induced per unit time, i.e., the light absorption rate, which in turn depends
on the available power intensity but also on the spectral characteristics of
the laser emission. Nevertheless, a large number of operating conditions
(pressure, cell dimensions, discharge intensity, gas purity, gas flow rate,
etc.) play a role in the OP process, simultaneously and in an intricate way.
It is thus very difficult to assess the efficiency of a particular laser
source in terms of absolute OP performances. One way out may be the comparison
of different types of laser sources under the same conditions \cite{Gentile02}%
. But the variety of experimental parameters that need to be controlled and
the difficulty encountered in obtaining identical beam divergences and
intensity profiles for lasers require considerable effort. The choice of
operating conditions may also be critical, either to emphasize the relevant
differences between the lasers or to test the sources in a configuration
suited for a particular application. A different approach is presented in this
article, which focuses on the laser spectral characteristics relevant for MEOP
of $^{3}\mathrm{He}$. The selected protocols provide conclusive results that
hardly depend on the experimental conditions. Detailed information on the
fiber laser line width and mode structure is obtained from purely optical
measurements. Further characterization with very basic equipment is provided
by tests of the atomic response on the $2^{3}\mathrm{S}$-$2^{3}\mathrm{P}$
transition at null nuclear polarization, which do not involve the intricate OP process.

\section{Architecture and operation of the fiber lasers\label{section3}}

\subsection{Tunable 1083 nm fiber lasers}

Different laser sources have been used to optically pump $^{3}\mathrm{He}$
gas: DBR semiconductor lasers \cite{Stoltz96,Major93}, arc lamp pumped Nd:LNA
lasers \cite{Daniels87}. Good results have been obtained with home-built
multimode LNA lasers \cite{Leduc90,Aminoff89,Gentile93}. Wide dissemination
and use outside the laboratory environment has, however, been hindered due to
inherent drawbacks and difficulties: modest provision and commercial
availability of high-quality crystal rods, limited output power (5-8~W, due to
severe thermal lensing effects inducing frequent rod breaking), low wall-plug
efficiency, and need for routine cleaning and re-alignment of the open-air
laser cavity optics. The emergence of fiber technology has allowed the
development of powerful light amplifiers (delivering tens of watts) and novel
laser sources in the near infrared region. Fiber lasers involving an external
cavity in Littrow configuration have recently been demonstrated to provide
good efficiency and wide tunability range \cite{Auerbach01}. However, as
explained in the previous section, narrowband emission sources are not well
suited for OP purposes.

Powerful broadband all-fiber lasers are therefore developed for OP of
$^{3}\mathrm{He}$\textrm{.} They are based on the convenient MOPFA
configuration. The master oscillator is a low power fiber laser designed for
tunable, multimode emission at 1083~nm with a spectral range on the order of
1-2~GHz. A large output power is provided by a power booster, which can be
separately optimized for maximal output efficiency. With adequate optical
isolation between the two parts and operation of the power amplifier in the
saturated regime, the spectral characteristics of the whole set-up are those
of the master oscillator. Both the master oscillator (characterized in this
work) and the power amplifier (described in detail in
\cite{BordaisTh,Bordais03}) take advantage of the high gain provided by
ytterbium-doped double clad fibers (DCFs) pumped by broad-stripe laser diodes.

A comprehensive introduction to the capabilities of Yb-doped fiber lasers can
be found, for instance, in \cite{Pask95}. The Yb$^{3+}$ ion advantageously
exhibits large absorption cross sections in the 915-925~nm range and a more
important one around 975~nm, where low-cost powerful laser diodes are
commercially available for pumping. In a Ge co-doped silica matrix,
transitions between sublevels of the ground $^{2}\mathrm{F}_{7/2}$ and excited
$^{2}\mathrm{F}_{5/2}$ states result in two main emission lines: a narrow
intense one at 975~nm and a broad one extending between 1 and 1.2~%
%TCIMACRO{\UNICODE{0xb5}}%
%BeginExpansion
$\mu$%
%EndExpansion
m. DCFs are preferred for high power operation, the large inner cladding
allowing propagation of the multimode pump beam around the doped active core
and multiple crossing across it. Continuous wave (CW) operation with output
powers exceeding 110~W has been demonstrated using DCF \cite{Dominic99}.

\subsection{The fiber oscillators}

The architecture of the fiber laser oscillator is based on a Fabry-P\'{e}rot
(FP) linear cavity closed by a high reflectivity fiber mirror and a tunable
fiber Bragg grating (FBG). The gain section consists of a Yb-doped DCF pumped
by a 975~nm laser diode (absorption: 0.9~dB/m) using the V-groove side pump
technique (VSP%
%TCIMACRO{\UNICODE{0xae}}%
%BeginExpansion
$^{\ooalign{\hfil\raise.07ex\hbox{$\scriptstyle\rm\text{R}$}\hfil\crcr\mathhexbox20D}}$%
%EndExpansion
: the pumping light is launched through imbedded V-grooves formed directly
into the DCF inner cladding and is totally reflected on the fiber side-wall,
yielding up to 90\% coupling efficiency \cite{Ripin85}). The doped fiber
length is about 3.5~m, it is not optimized in terms of output power and pump
absorption in this work. Inside the protective polymer outer cladding, the
inner glass cladding is about 200~%
%TCIMACRO{\UNICODE{0xb5}}%
%BeginExpansion
$\mu$%
%EndExpansion
m in diameter with a numerical aperture of 0.45 and a star shape profile to
ensure efficient mode mixing and high absorption of the pump light in the
doped area. A pair of all-glass silica fibers, single mode at 1~%
%TCIMACRO{\UNICODE{0xb5}}%
%BeginExpansion
$\mu$%
%EndExpansion
m, are spliced at both ends of the DCF to allow connection to the other fiber
optical components.

The tunable FBG, used for frequency discrimination, serves as a low
transmission output mirror for the laser cavity. The two key characteristics
of the FBG are thus the FWHM of its operating range and the reflectivity. They
both play an important role in setting the laser power and line width, and
must be optimized and controlled. The tunable FBG is thermally isolated and
temperature regulated to stabilize the laser frequency. Piezoelectrical
control of mechanical strain applied to the grating allows fine-tuning of the
laser frequency to the helium resonance lines.

\begin{figure}[th]
%h=here, t=top, b=bottom, p=separate figure page
%
%
%
%
%
%
%
%
%
%
%
%
%
%
%
%
%
%
%
%
%
%
%
%
%
%
%
%
%
%
%
%
%
%
%
%
%
%
%
%
%
%
%
%
%
%
%
%
%
\par
\begin{center}
\leavevmode
\includegraphics[keepaspectratio,width=0.95\linewidth,clip= ]{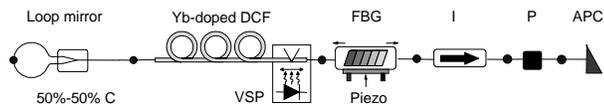}
\end{center}
\caption{1083~nm tunable broadband fiber oscillator. The linear laser cavity
includes a fiber loop mirror formed by splicing the two ends of a 50\%-50\%
coupler (C), a Yb-doped DCF and a selective FBG with piezo-control. An
isolator (I) and a fiber polarizer (P) are fusion-spliced (solid dots) before
the output angle polished connector (APC). V-groove side pumping (VSP) of the
fiber is performed at 975~nm.}%
\label{figure2}%
\end{figure}

Fig.~\ref{figure2} displays the architecture of a fiber oscillator in which a
loop mirror is used to close the FP linear cavity. Fiber loop reflectors are easily fabricated and exhibit large
reflectivities (50-75\%) over a broad frequency range \cite{Mortimore88}. The
selective tunable FBG is spliced at the output end of the cavity. It directly
controls both the operation frequency and the FWHM of the laser.

The operation of that loop mirror oscillator is hereafter fully characterized.
Subsequent comparison is made with another fiber oscillator based on the same
FP linear cavity architecture, where the loop mirror is replaced by a high
reflectivity FBG to fully optimize the laser for OP applications (see
Fig.~\ref{figure12} in Sect.~\ref{section4}, that depicts both this second
oscillator and the power amplifier used to build the complete all-fiber MOPFA
laser). The two fiber oscillators mainly differ by their temporal and spectral behavior.

\subsection{Operation of the loop mirror oscillator}

\begin{figure}[th]
%h=here, t=top, b=bottom, p=separate figure page
%
%
%
%
%
%
%
%
%
%
%
%
%
%
%
%
%
%
%
%
%
%
%
%
%
%
%
%
%
%
%
%
%
%
%
%
%
%
%
%
%
%
%
%
%
%
%
%
%
\par
\begin{center}
\leavevmode
\includegraphics[keepaspectratio,width=0.95\linewidth,clip= ]{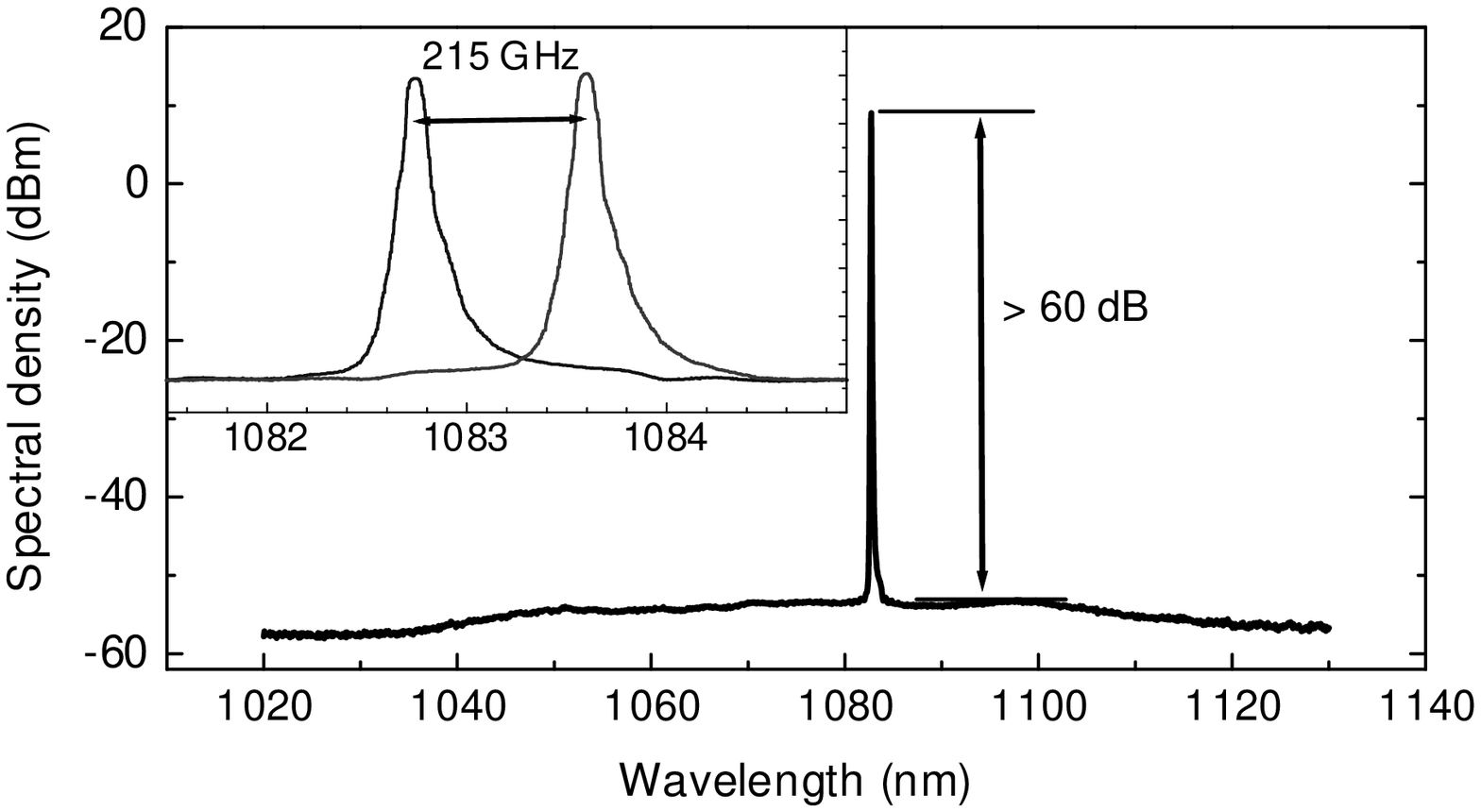}
\end{center}
\caption{Optical spectrum of the loop mirror oscillator (resolution: 0.07~nm).
Insert: fine tuning to the $\mathrm{He}$ lines is achieved by piezo-control of
the selective FBG.}%
\label{figure3}%
\end{figure}

The optical spectrum of the loop mirror oscillator is shown in
Fig.~\ref{figure3}. The emission range of the oscillator duly meets the laser
requirements. It includes the whole $^{3}\mathrm{He}$ and $^{4}\mathrm{He}$
absorption spectrum (0.275~nm broad, starting at 1082.908~nm in air for
D$_{0}$ at null magnetic field). Piezo-control of the FBG actually allows fine
tuning of the laser over 0.8~nm, i.e., more than 200~GHz (Fig.~\ref{figure3},
insert). Therefore, all components of the $2^{3}\mathrm{S}$-$2^{3}\mathrm{P}$
transition can be reached, even for OP in a 1.5~T applied magnetic field as
indicated in Sect.~\ref{section2}.

\begin{figure}[th]
%h=here, t=top, b=bottom, p=separate figure page
%
%
%
%
%
%
%
%
%
%
%
%
%
%
%
%
%
%
%
%
%
%
%
%
%
%
%
%
%
%
%
%
%
%
%
%
%
%
%
%
%
%
%
%
%
%
%
%
%
\par
\begin{center}
\leavevmode
\includegraphics[keepaspectratio,width=0.95\linewidth,clip= ]{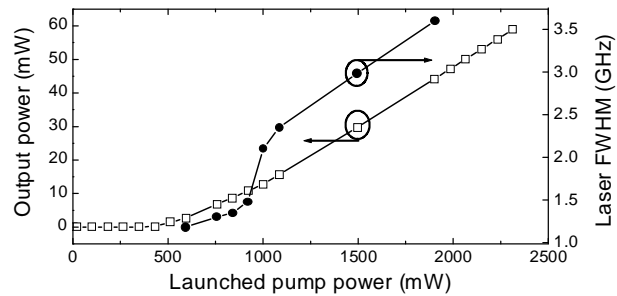}
\end{center}
\caption{Output power (open squares, left vertical axis) and bandwidth (solid
dots, right vertical axis) measured for the loop mirror oscillator. An abrupt
FWHM increase occurs above $P_{\mathrm{pump}}$=920~mW. }%
\label{figure4}%
\end{figure}

The loop mirror oscillator delivers up to 60~mW for a launched pump power
$P_{\mathrm{pump}}$=1900~mW, with a laser threshold around $P_{\mathrm{pump}}%
$=500~mW (Fig.~\ref{figure4}, open squares). A polarization-independent isolator protects the oscillator from
accidentally back-reflected light during operation. Linear polarization is
achieved by a fiber polarizer spliced at the end of the laser cavity
(extinction ratio
%TCIMACRO{\TEXTsymbol{>}}%
%BeginExpansion
$>$%
%EndExpansion
20~dB). These two components induce a total 3 dB loss, taken into account to
determine the power actually delivered at the output of the tunable FBG in
Fig.~\ref{figure4}.

The average spectral bandwidth (Fig.~\ref{figure4}, solid dots) is measured
using a FP analyzer (an air gap etalon whose cavity length is piezomodulated
at low frequency) with typically 10~GHz free spectral range (FSR) and 80
finesse. The laser modes are not resolved, the laser cavity FSR being in the
MHz range. Substantial FWHM broadening is observed when the pump power is
increased. An abrupt change in FWHM occurs above $P_{\mathrm{pump}}$=920~mW,
in the absence of any correlated modification of the laser output power.

\subsection{Temporal behavior and mode structure}

\begin{figure}[th]
%h=here, t=top, b=bottom, p=separate figure page
%
%
%
%
%
%
%
%
%
%
%
%
%
%
%
%
%
%
%
%
%
%
%
%
%
%
%
%
%
%
%
%
%
%
%
%
%
%
%
%
%
%
%
%
%
%
%
%
%
\par
\begin{center}
\leavevmode
\includegraphics[keepaspectratio,width=0.95\linewidth,clip= ]{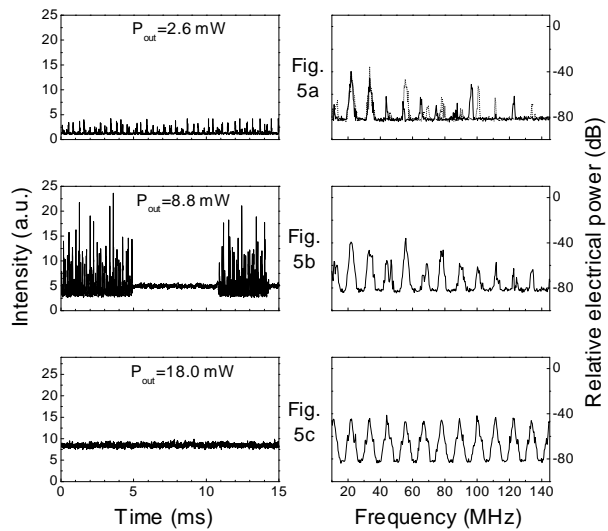}
\end{center}
\caption{Time variations (left graphs) and frequency spectra (right graphs) of
the laser intensity for the loop mirror oscillator. Fig.~5a:~$P_{\mathrm{pump}%
}$ = 593~mW; self-pulsing regime, with only a few emitted modes. The dotted
line added in the right graph is a spectrum obtained from another randomly
triggered singleshot recording (see text). Fig.~5b:~$P_{\mathrm{pump}}$ =
839~mW; unstable regime, switching between pulsed and CW operation.
Fig.~5c:~$P_{\mathrm{pump}}$= 1167~mW; stable regime, with CW and highly
multimode operation. The output powers of the oscillator are indicated in the
left graphs ($P_{\mathrm{out}}$).}%
\label{figure5}%
\end{figure}

Intensity noise (IN) measurements are performed for the loop mirror oscillator
described in Fig.~\ref{figure2} using a photodiode with 700~MHz bandwidth
connected both to a fast digital oscilloscope and\ to an electrical spectrum
analyzer (ESA). Fig.~\ref{figure5} displays time variations of the output
intensity and frequency spectra obtained at three pump powers
($P_{\mathrm{pump}}$ = 593, 839 and 1167~mW). The frequency resolution (1~MHz) does not allow to characterize
the frequency and repetition rate of the relaxation oscillations in the
self-pulsing regime (see below and \cite{BordaisTh,Bordais03,Hideur00}). The
peaks seen on the ESA output traces result from longitudinal modes emitted
under the laser line width envelope beating together. This inter-modal beat
signal gives access to the number and frequency splitting of the modes emitted
on the probed time scale.

Three main types of behavior can be distinguished as the pump power is
increased. The first one is observed near laser threshold (e.g.,
Fig.~\ref{figure5}a: $P_{\mathrm{pump}}$=593~mW, laser FWHM 1.2~GHz). The
laser operates spontaneously in self-pulsing regime, delivering series of
pulses of sometimes relatively high peak power. This behavior is thus
potentially harmful in view of subsequent high-power amplification. The
corresponding IN spectra only include beat frequencies\ below 150~MHz. Their
statistical analysis shows that only peaks up to 40~MHz are always present, a
few extra ones randomly appearing at frequencies that are other integer
multiple of the FSR of the laser cavity. The second type of behavior (e.g.,
Fig.~\ref{figure5}b: $P_{\mathrm{pump}}$=839~mW, laser FWHM still below
1.5~GHz) is characterized by emission of intense light pulses alternating in
time with CW light emission. In this case, inter-modal beat frequencies are
regularly and permanently distributed under a typically 150~MHz wide envelope.
The third and last type of behavior (e.g., Fig.~\ref{figure5}c:
$P_{\mathrm{pump}}$=1167~mW and above) corresponds to a stable regime with a
nonzero power continuously emitted, even on atomic lifetime scale (0.1~$\mu
$s). A very large number of modes simultaneously exist inside the cavity$:$
the IN\ spectrum is a comb of peaks with uniform frequency splitting (the
cavity FSR) extending over the whole instrument limited detection bandwidth
(the laser FWMH has jumped over the sharp discontinuity seen in
Fig.~\ref{figure4} and is in that regime significantly larger than 2~GHz).

Strong non linear effects are known to occur in high-gain rare-earth-doped
fibers. The high powers supported by the Yb-doped DCF can lead to
instabilities or huge intensity fluctuations due to distributed backscattering
effects that are exacerbated in the linear cavity configuration
\cite{Chernikov97b,Hideur00}. Substantial variations of the laser bandwidth
with pump power have been previously reported \cite{Auerbach02}, and intensity
fluctuations as well as changes of the mode structure investigated
\cite{Hideur00}, for Yb-doped fiber lasers in Littrow configuration. These
dynamical behaviors are reminiscent of those obtained in Er-doped fiber lasers
\cite{Lacot91,Sanchez93}, but their exact physical origin in Yb-doped fibers
may not be yet fully established \cite{BordaisTh,Bordais03,Pask95,Hideur00}.
Self-pulsing has been demonstrated to be dependent on passive cavity losses
and favored at high losses \cite{BordaisTh,Bordais03,Ortac03}. The loop mirror
used in the oscillator cavity has an intrinsic loss of order 1~dB and the
reflectivity of the FBG is optimized. However significant losses may be due to
technical difficulties encountered in achieving quality splices between the
Yb-doped DCF and the pure silica fiber used for the junctions.
Polarization-dependent losses induced by the loop mirror are also expected to
contribute to the occurrence of the self-pulsing regime. With the loop mirror
oscillator, similar $P_{\mathrm{pump}}$ threshold values are observed for the
changes of mode structure and temporal dynamics and for the variations of
average FWHM. However the comparison of different types of linear cavity
lasers indicates that it may well be fortuitous \cite{BordaisTh,Bordais03}.

\begin{figure}[th]
%h=here, t=top, b=bottom, p=separate figure page
%
%
%
%
%
%
%
%
%
%
%
%
%
%
%
%
%
%
%
%
%
%
%
%
%
%
%
%
%
%
%
%
%
%
%
%
%
%
%
%
%
%
%
%
%
%
%
%
%
\par
\begin{center}
\leavevmode
\includegraphics[keepaspectratio,width=0.95\linewidth,clip= ]{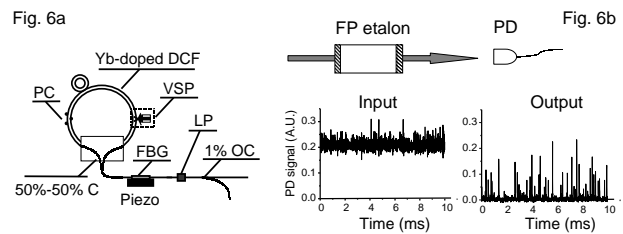}
\end{center}
\caption{Fig.~6a: Ring cavity 1083~nm fiber oscillator (1.64~GHz FWHM) tested
in \cite{CourtadeTh}. PC: polarization controller, LP: linear polarizer, (O)C:
(output) coupler. Fig.~6b, top part: Experimental setup used for time-resolved
spectral analysis, including a FP etalon (1.5~GHz FSR, finesse: 100) and a
fast photodiode (PD). Fig.~6b, bottom graph: Time variations of the input and
output PD signalobtained, respectively, without and with the FP etalon along
the laser beam (solid arrow in top part of Fig.~6b).}%
\label{figure6}%
\end{figure}

It is worth noting the correlation observed for this loop mirror oscillator
between temporal behavior and mode structure: self-pulsing behavior and a
limited number of emitted modes (Fig.~\ref{figure5}a), or stable intensity and
multimode operation (Fig.~\ref{figure5}c). In contrast, a different fiber
oscillator based on a ring laser cavity (Fig.~\ref{figure6}) exhibited stable
output power but quasisinglemode structure \cite{CourtadeTh}.
For that ring cavity laser (1.64~GHz FWHM) no ESA measurements had
been performed. But on one hand the laser intensity had been monitored using a
fast photodiode (response time well below 1~$\mu$s) and was indeed stable in
time. On the other hand the light transmitted through a FP etalon with fixed
length, FSR 1.5~GHz and spectral bandwidth 15~MHz (i.e., comparable to the
$2^{3}\mathrm{S}$-$2^{3}\mathrm{P}$ radiative line width in the presence of
collisions in a 1~mbar cell), was discontinuous. Nonzero transmitted
intensities could be observed only for very short periods of time (6.7~$\mu$s
on average, with 0.2~$\mu$s rise time and 0.2-0.5~$\mu$s fall time at 1/$e$),
separated by long periods of null transmission (215~$\mu$s long on average).
Assuming a large mode jitter over the observed 1.64~GHz FWHM to result in
random coincidence of the modes frequencies with the transmission bandwidth of
the FP etalon, statistical analysis of repeated measurements lead to the
conclusion that only 3 or 4 laser modes were simultaneously emitted. This had
been confirmed later by a low saturation threshold in light absorption
measurements (method and meaning described in section~4) and explains the
moderate OP\ performances of this ring laser despite its proper FWHM and CW
operation \cite{CourtadeTh}.

Therefore the same FP filter is used with the loop mirror oscillator to
perform similar tests, for comparison and cross-check with the ESA
measurements. CW light is indeed transmitted above $P_{\mathrm{pump}}\simeq
$1300~mW, suggesting truly multimode laser operation on the relevant atomic
time scale (0.1~$\mu$s) in the stable regime. In contrast, in the self-pulsing
regime light is seldom emitted in the FP spectral transmission bandwidth and
the average durations of the null-transmission periods typically vary from
1~ms close to laser threshold, to 0.1~ms at $P_{\mathrm{pump}}$=760~mW for
instance. These observations fully confirm the conclusions drawn from the
above-described IN measurements.

\subsection{Mode dynamics}

The comparison between the IN spectra and the measured FWHM of the loop mirror
oscillator suggests that in the self-pulsing regime the measured broad laser
bandwidth results from very large mode jitter inside the cavity. In order to
confirm this and to directly observe the mode dynamics, we observe the time
variation of the beat signal between the fiber oscillator output and that of a
singlemode 1083~nm DBR diode (50~mW output power, 3~MHz bandwidth
\cite{Stoltz96}). To conveniently tune the two lasers to one another to better
than 1~GHz, a fluorescence signal from a sealed $^{3}\mathrm{He}$ cell is used.

\begin{figure}[th]
%h=here, t=top, b=bottom, p=separate figure page
%
%
%
%
%
%
%
%
%
%
%
%
%
%
%
%
%
%
%
%
%
%
%
%
%
%
%
%
%
%
%
%
%
%
%
%
%
%
%
%
%
%
%
%
%
%
%
%
%
\par
\begin{center}
\leavevmode
\includegraphics[keepaspectratio,width=0.95\linewidth,clip= ]{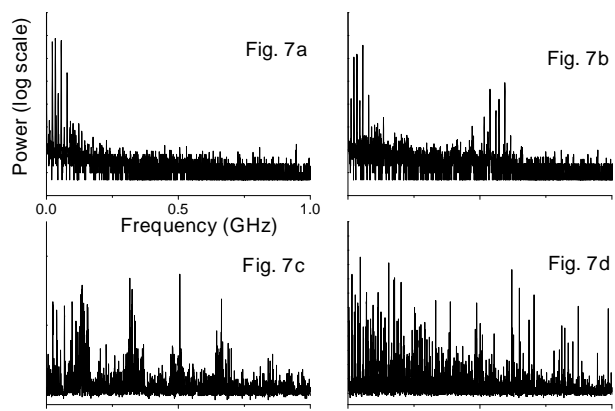}
\end{center}
\caption{IN\ spectra of the loop mirror oscillator beam alone (Fig.~7a) or
superimposed with that of a DBR\ diode (Fig.~7b to 7d), in the self-pulsing
regime ($P_{\mathrm{pump}}$ = 593~mW). All plots have identical frequency
span, 1~GHz full scale. Fig.~7a and 7b: 0.1~$\mu$s recording time. The
autocorrelation spectrum of the oscillator (Fig.~7a) also appears in Fig.~7b
at low frequencies. The beat signal of the fiber laser and the singlemode
diode (set of peaks around 0.6~GHz in Fig.~7b) indicates that only a few modes
are simultaneously emitted. Mode jitter is evidenced by the broad spread of
the beat peak frequencies as the recording time increases (Fig.~7c: 5~ms,
Fig.~7d: 0.1~s).}%
\label{figure7}%
\end{figure}

Fig.~\ref{figure7}b to \ref{figure7}d display power spectra obtained by fast
Fourier transform of single-shot digital recordings (duration: 0.1~$\mu$s,
5~ms and 0.1~s, respectively) of the signal delivered by a fast photodiode
located on the path of the two superimposed beams. The DBR diode is operating at fixed frequency (typically a few
hundred MHz away from the central frequency of the fiber oscillator) and the
oscillator is used at low pump power ($P_{\mathrm{pump}}$ = 593~mW,
self-pulsing regime). For comparison, a spectrum obtained without the DBR
diode beam (recording time: 0.1~$\mu$s) is shown in Fig.~\ref{figure7}a. The
low frequency parts of the heterodyne IN spectra (DBR diode plus fiber
oscillator) in Fig.~\ref{figure7}b to \ref{figure7}d are identical to those
obtained with the fiber oscillator alone (homodyne autocorrelation spectrum)
using the same set-up (Fig.~\ref{figure7}a) or using the ESA (see
Fig.~\ref{figure5}a). As detailed in the previous subsection, they correspond
to the intermodal beat signal of the fiber oscillator. At higher frequencies,
series of peaks appear which correspond to the beat signal of the diode and
the oscillator. Frequency splittings are here again set by the fiber cavity
FSR. Only one comb of beat peaks appears at very short recording times
(Fig.~\ref{figure7}b). Repeated measurements demonstrate that close to laser
threshold a very small number of contiguous modes are indeed emitted on the
atomic time scale (0.1~$\mu$s). On one hand the central frequency of this comb
of beat peaks is observed to change from shot to shot, randomly located within
the detection bandwidth (of order 1~GHz). The frequency stability of the DBR
diode is checked by the observation of the beat signal between two different
DBR diodes in the same experimental conditions. The time variations of the
beat frequencies are thus attributable to the oscillator only. On the other
hand the spectra of single-shot recordings taken on much longer time scales
(several ms or more) display a continuous comb of beat peaks extending beyond
1~GHz (see Fig.~\ref{figure7}d). Both observations confirm that a large jitter
of a small number modes is responsible for the large average oscillator FWHM
measured with the modulated FP etalon in the self-pulsing regime. They also
indicate that for He atoms the laser would appear as quasisinglemode despite
this broad apparent optical spectrum.

Away from the laser threshold, the number of modes beating with the diode
increases with the pump power. Broadband ($\geq$1~GHz FWHM) multimode
operation on the atomic time scale is observed in the stable regime. However
no further information can be obtained at high $P_{pump}$ due to the limited
detection bandwidth in this experiment.

Finally the observed temporal behavior, mode structure and dynamics are
checked to remain totally unchanged when the output of the laser oscillator is
launched into a saturated fiber power amplifier, as expected.

\section{Atomic response to laser characteristics\label{section4}}

\subsection{Line width measurements}

Unsophisticated but accurate measurements of the laser average FWHM can be
achieved using a sample of $^{3}\mathrm{He}$ gas. The piezocontrol is used to
sweep the oscillator frequency over the C$_{8}$ and C$_{9}$ lines.
Fluorescence light (or laser absorption) at null nuclear polarization is
monitored using a photodiode to extract the resonance line width for a weak
probe beam intensity. Power densities well below a few mW/cm$^{2}$ are needed
to avoid saturation (see next subsection). When circularly polarized light is
used, a strong local field inhomogeneity must be imposed in the He cell to
prevent nuclear polarization build-up, which would significantly alter the
line profiles \cite{Courtade02}. At high laser intensities finite orientation
can still be produced in the metastable state despite strong nuclear
relaxation in the ground state (over-polarization of the $2^{3}\mathrm{S}$
atoms \cite{Leduc00}) and the best way to avoid OP effects is to use linearly
polarized light. In any case, the measured relative amplitudes of the two
resonance peaks must correspond to the known ratio of the C$_{8}$ and C$_{9}$
line intensities (1.284 at very low magnetic field). Absolute calibration of
the frequency scale is provided by the hyperfine splitting between C$_{8}$ and
C$_{9}$ (6.74~GHz). The experimental resonance line width results from the
convolution of the atomic Doppler profile with the spectral envelope of the laser.

\begin{figure}[th]
%h=here, t=top, b=bottom, p=separate figure page
%
%
%
%
%
%
%
%
%
%
%
%
%
%
%
%
%
%
%
%
%
%
%
%
%
%
%
%
%
%
%
%
%
%
%
%
%
%
%
%
%
%
%
%
%
%
%
%
%
\par
\begin{center}
\leavevmode
\includegraphics[keepaspectratio,width=0.95\linewidth,clip= ]{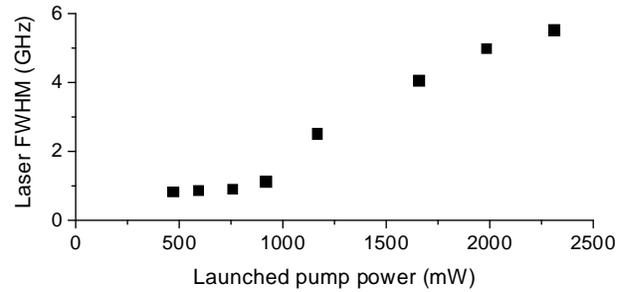}
\end{center}
\caption{FWHM\ of the loop mirror oscillator obtained from C$_{8}$ and C$_{9}$
resonance linewidth measurements.}%
\label{figure8}%
\end{figure}

Recorded signals are very well fit by a set of two Gaussians of FWHM
$D_{\mathrm{exp}}$, which indicates that the laser spectral profile is
actually very close to Gaussian \cite{BordaisTh,Bordais03}.\ The laser FWHM
$L$ is thus given by $L^{2}=D_{\mathrm{exp}}^{2}-D^{2}$. Fig.~\ref{figure8}
displays the obtained laser FWHMs as a function of the pump power for the loop
mirror oscillator. The measured variation is in good qualitative agreement with the
behavior discussed in section~\ref{section3} (little change up to
$P_{\mathrm{pump}}$=920~mW, then abrupt broadening and linear increase above
$P_{\mathrm{pump}}$=1100~mW). The noticeable quantitative difference
(20-30\%)\ between these FWHM\ results and those obtained with the FP analyzer
(Fig.~\ref{figure4}, solid dots) mainly arises from the fact that two
measurements were not performed in the same frequency range. Indeed, over the
200~GHz broad tuning range, the operating conditions of the laser cavity vary
due to changes in the loop mirror reflectivity \cite{Mortimore88}. The
influence of the mirror characteristics on the laser FWHM is analyzed and
discussed in more detail in \cite{BordaisTh,Bordais03}. Experimentally, using
the piezomodulated FP analyzer to monitor the laser FWHM (see
section~\ref{section3}), we actually observe changes of order 20\% with
operating frequency between the extreme limits of the tuning range of the loop
mirror oscillator. The amplitudes of these FWHM\ changes depend on how close
to threshold the laser is operated. Therefore, for OP applications the laser
should be kept working in stable operating conditions and precisely
characterized at the atomic frequency used to polarize the gas.

\subsection{Absorption measurements}

Absorption measurements provide quantitative information on the actual
fraction of light that can be used for efficient OP of the atomic nuclei.
Laser absorption is measured as a function of the incident power density for
the oscillator tuned to the C$_{9}$ resonance line, in a sealed cell (5~cm
long, 5~cm in diameter) filled with 1.6~mbar of ultrapure $^{3}\mathrm{He}$
gas. In order to perform accurate comparison between several operating
conditions of the laser cavity measurements are performed successively with
identical discharge conditions to keep the $2^{3}\mathrm{S}$ number density
constant. The same basic equipment is required, except that a lock-in
amplifier advantageously improves the signal-to-noise ratio at low laser
powers. As the laser power is varied, care is taken to keep the photodiode
operating at nearly constant output level (using attenuators) to avoid
possible biases due to saturation or non-linear response of the device. The
gas is also always maintained at null nuclear polarization. The laser beam
with Gaussian transverse intensity profile is expanded and a diaphragm of
known open area is inserted to select a small central fraction of the
collimated beam. Over the probed section the radial variation of the laser
power has been checked not to exceed a few percents, so that the
(quasi)uniform power densities are precisely determined from absolute power
measurements. To reach strong optical saturation of the gas sample and to
complete the investigations close to laser threshold, a 0.5~W fiber amplifier
is used to boost up the light power delivered by the laser oscillator.

\begin{figure}[th]
%h=here, t=top, b=bottom, p=separate figure page
%
%
%
%
%
%
%
%
%
%
%
%
%
%
%
%
%
%
%
%
%
%
%
%
%
%
%
%
%
%
%
%
%
%
%
%
%
%
%
%
%
%
%
%
%
%
%
%
%
\par
\begin{center}
\leavevmode
\includegraphics[keepaspectratio,width=0.95\linewidth,clip= ]{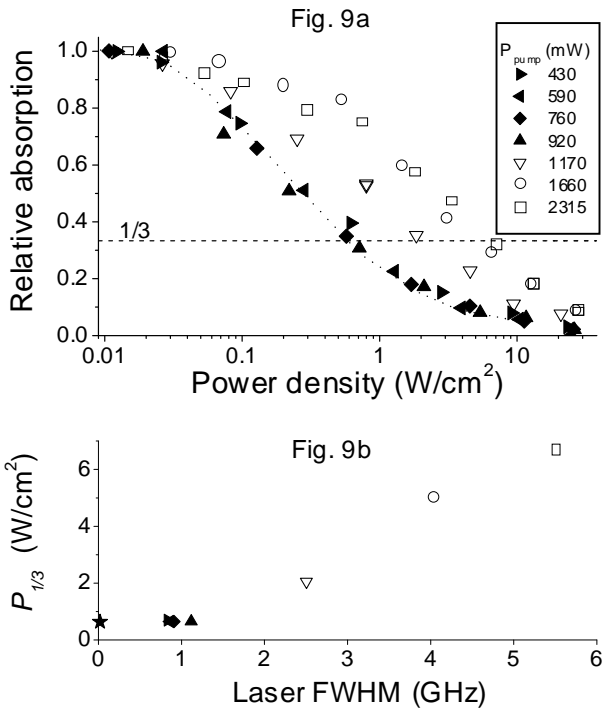}
\end{center}
\caption{Fig.~\ref{figure9}a: Relative absorption (see text) versus laser
power density. Symbols: loop mirror oscillator operated at low (solid symbols)
and high (open symbols) launched pump powers; 1.6~mbar $^{3}\mathrm{He}$ cell,
light with circular polarization tuned on the C$_{9}$ line. Dotted line:
singlemode DBR diode, reference absorption profile from \cite{CourtadeTh};
same cell, very similar experimental conditions. Fig.~\ref{figure9}b:
Saturation threshold $\mathcal{P}_{\mathrm{1/3}}$ (see text) versus laser
FWHM. Solid and open symbols from Fig.~\ref{figure9}a: loop mirror oscillator.
Star: singlemode DBR diode.}%
\label{figure9}%
\end{figure}

The relative absorption data plotted in Fig.~\ref{figure9}a correspond to
absorption rates measured with the loop mirror oscillator for different values
of the launched pump power, normalized to the maximum absorption rate obtained
at very low power intensities for each $P_{\mathrm{pump}}$. At low launched pump powers (solid symbols) the absorption
profiles do not vary much with $P_{\mathrm{pump}}$. They are quite similar to
those\ usually obtained with a singlemode DBR diode (Fig.~\ref{figure9}a,
dotted line: data from reference \cite{CourtadeTh}, obtained in the same cell
with comparable discharge conditions and measurement protocol). This confirms
the conclusion drawn from the optical measurements and spectral analysis in
section~\ref{section3} for the unstable and self-pulsing regimes. As pump
power is increased, the strong decrease of absorption due to optical
saturation suddenly occurs at much higher power densities. This change in
atomic response to laser excitation is a signature of the abrupt transition of
the oscillator to the broadband multimode emission regime. At high launched
pump powers (above $P_{\mathrm{pump}}$=920~mW, open symbols) the relative
absorptions measured for a fixed power density become significantly larger
when $P_{\mathrm{pump}}$ is increased

To characterize this change in behavior we introduce the power density
$\mathcal{P}_{\mathrm{1/3}}$ where the measured absorption reaches 1/3 of its
maximum value (dashed line in Fig.~\ref{figure9}a). $\mathcal{P}%
_{\mathrm{1/3}}$ is plotted in Fig.~\ref{figure9}b as a function of the loop
mirror oscillator FWHM. It indeed remains almost constant up to
$P_{\mathrm{pump}}$=920~mW, and close to the 0.64~W/cm$^{2}$ value typically
obtained with the singlemode diode (solid star, Fig.~\ref{figure9}b). It then
increases quite linearly with the laser FWHM above $P_{\mathrm{pump}}%
$=1100~mW. This sudden change of absorption saturation threshold correlates
precisely with the abrupt change in spectral behavior described in
section~\ref{section3}.

\begin{figure}[th]
%h=here, t=top, b=bottom, p=separate figure page
%
%
%
%
%
%
%
%
%
%
%
%
%
%
%
%
%
%
%
%
%
%
%
%
%
%
%
%
%
%
%
%
%
%
%
%
%
%
%
%
%
%
%
%
%
%
%
%
%
\par
\begin{center}
\leavevmode
\includegraphics[keepaspectratio,width=0.95\linewidth,clip= ]{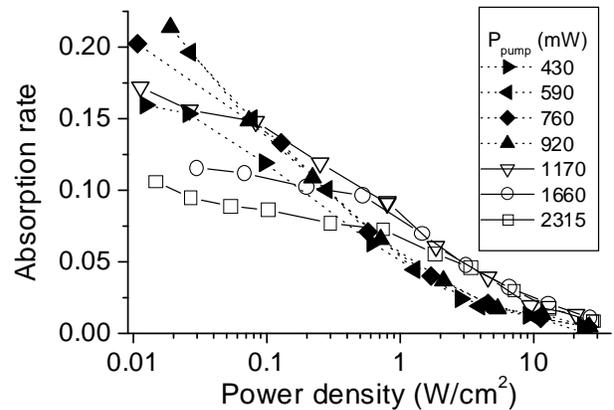}
\end{center}
\caption{Absorption rates versus power densities for the loop mirror
oscillator (same data sets as in Fig.~\ref{figure9}, but without normalization
to maximum values). The variations of absorption rates, hence of laser
efficiency, with launched pump powers result from changes in the number of
emitted laser modes (despite identical average FWHM, solid symbols) or from
changes in the amount of off-resonance laser power (open symbols).}%
\label{figure10}%
\end{figure}

A relevant quantity for the OP process is the total number of absorbed photons
per second, since it is the angular momentum deposition rate that determines
the actual number of oriented nuclei. Absolute absorption rates are therefore
plotted in Fig.~\ref{figure10} (same measurements performed with the loop
mirror oscillator, data without normalization). At all light power densities the absorption rates are observed to
vary by more than a factor 2 over the probed range of launched pump powers.
For small $P_{\mathrm{pump}}$ (solid symbols) the absorption rates at low
laser intensities rapidly vary with power density down to 10~mW/cm$^{2}.$ They
tend to increase with $P_{\mathrm{pump}}$, as can be expected due to the
increased number of emitted laser modes at resonance with the atoms.
Differences, however, become hardly noticeable above a few hundreds
mW/cm$^{2}$. For $P_{\mathrm{pump}}>$920~mW (open symbols) saturation effects
are small at very and moderately low power densities. The maximum absorption
rates decrease when $P_{\mathrm{pump}}$ rises due to excessively large laser
FWHMs (useless power emitted outside the Doppler line). At higher laser powers
(above 1~W/cm$^{2}$) the absorption rate is nearly twice larger than that
obtained at low $P_{\mathrm{pump}}$ because of multimode light emission by the
loop mirror oscillator. This absorption rate is observed to be independent of
$P_{\mathrm{pump}}$, which means that the measured change of relative
absorption and saturation threshold (Fig.~\ref{figure9}) is exactly
compensated by the effect of strongly increased FWHMs in that regime
(Fig.~\ref{figure8}). Above 20~W/cm$^{2}$ very strong optical saturation
occurs regardless of the laser FWHM and operation regime (absortion rates well
below 1\%).

The observed modification of absorption profiles and atomic resonance line
saturation rates is evidence of a strong variation of the spectral coverage
between the Doppler and laser profiles under the influence of the operating
conditions of the loop mirror cavity. The absorption measurements do not
provide as many details on the mode structure and behavior of the fiber laser
as do the optical spectral measurements in section~\ref{section3}. But they
are rather easy to carry out, and the experimental data can be quantitatively
analyzed to get a relevant characterization of the laser features for OP\ purposes.

\begin{figure}[th]
%h=here, t=top, b=bottom, p=separate figure page
%
%
%
%
%
%
%
%
%
%
%
%
%
%
%
%
%
%
%
%
%
%
%
%
%
%
%
%
%
%
%
%
%
%
%
%
%
%
%
%
%
%
%
%
%
%
%
%
%
\par
\begin{center}
\leavevmode
\includegraphics[keepaspectratio,width=0.95\linewidth,clip= ]{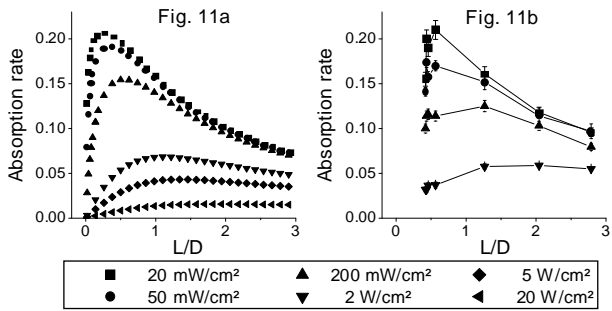}
\end{center}
\caption{Absorption versus laser FWHM $L$ (scaled to the Doppler width $D$),
for various light power densities. Fig. 11a: Computed results using the
phenomenological OP model \cite{Leduc00}. Fig. 11b: Experimental measurements
with the loop mirror oscillator (data sets extracted from Fig.~\ref{figure10}%
).}%
\label{figure11}%
\end{figure}

To emphasize the influence of the laser spectral characteristics on the
response of the atomic system, Fig.~\ref{figure11} displays the change of
absorption rates with the laser FWHM for various light power densities shined
onto the gas sample. Fig.~\ref{figure11}a shows the expected crossover behavior between
the high and low power limits qualitatively discussed above.\ These results
are obtained with the phenomenological OP model described in \cite{Leduc00},
using for the spectral coverage parameter $X_{\mathrm{s}}$ the rough estimate
$erf(\sqrt{Ln(2)}L/D)$ introduced in section~\ref{section2}. The experimental
data are plotted in Fig.~\ref{figure11}b. Measurements appear to be in good
qualitative agreement with the computed estimates. For $P_{\mathrm{pump}}%
>$920~mW (data points with $L/D\geq1$ in Fig.~\ref{figure11}b) we observe at
the lowest power densities the expected strong absorption decrease with laser
FWHM, above a maximum located somewhere below $L/D=$~0.5. The dominant
contribution here comes from the growing fraction of off resonance photons in
the incident beam, resulting in the $1/\sqrt{1+L^{2}/D^{2}}$ scaling discussed
in section~\ref{section2}. The location of the absorption maximum moves
towards $L/D\geq1$ as the laser power increases as indicated by the
computations. At very high powers absorption stops varying with the laser FWHM
for $L/D>1$. Relevant experimental data with $L/D<1$ are not available with
the loop mirror oscillator. The results obtained outside the multimode regime
(appearing at $L/D<0.56$ in Fig.~\ref{figure11}b) have been included only to
visualize the amplitude of absorption changes observed at small
$P_{\mathrm{pump}}$. But, as demonstrated in the previous section, the
measured FWHMs (1-1.1~GHz below $P_{\mathrm{pump}}<$~920~mW) indeed result
from frequency jitter and on their time scale the atoms actually interact with
a finite set of discrete modes. This situation might tentatively be described
by a set narrow continuous distributions with increasing effective FWHMs (to
account for the observed change in IN spectra envelopes with $P_{\mathrm{pump}%
}$). This would certainly make the low $P_{\mathrm{pump}}$ data delineate
clear absorption maxima in Fig.~\ref{figure11}b, with location and rounded
shape closer to expectations. However, the values of these effective FWHMs are
difficult to accurately assess a priori, and their meaning and relevance for
further studies are questionable considering the crude description of
interaction probabilities between atoms and photons in the model.

\subsection{Laser optimization for OP applications}

Data in Fig.~\ref{figure10} and \ref{figure11} indicate that, to optimize
absorption, the laser FWHM should lie somewhere between 1 and 2.4~GHz,
depending on laser intensity. For OP purposes the laser intensities are
usually on the order of 1~W/cm$^{2}$. Therefore the optimal emission bandwidth
should indeed be comparable to the atomic Doppler line width, as qualitatively
expected from the discussion in section~\ref{section2}. However, this FWHM
range cannot be properly investigated with the loop mirror cavity oscillator,
due to the abrupt FWHM\ jump from 1.1 to 2.5~GHz (see Fig.~\ref{figure8}) as
well as the switching from quasi-singlemode to truly multimode operation. The
cavity configuration needs to be changed to achieve multimode operation and
laser FWHM adjustable down to 1~GHz.

\begin{figure}[bh]
%h=here, t=top, b=bottom, p=separate figure page
%
%
%
%
%
%
%
%
%
%
%
%
%
%
%
%
%
%
%
%
%
%
%
%
%
%
%
%
%
%
%
%
%
%
%
%
%
%
%
%
%
%
%
%
%
%
%
%
%
\par
\begin{center}
\leavevmode
\includegraphics[keepaspectratio,width=0.95\linewidth,clip= ]{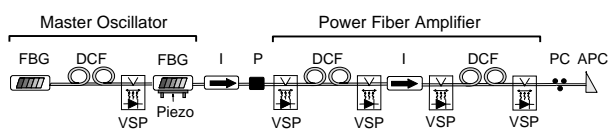}
\end{center}
\caption{All-fiber 1083~nm MOPFA laser optimized for MEOP applications. The
dedicated fiber oscillator with double FBG linear cavity is attached to a
two-stage Yb-doped DCF power amplifier. VSP: V-groove side pumping; I: optical
isolator; P: polarizer; PC: polarization controller; APC: angle polished
connector. }%
\label{figure12}%
\end{figure}

This is achieved with double FBG linear cavities, using a high reflectivity
fiber Bragg grating instead of the loop mirror. The architecture of the double
FBG oscillator is depicted in Fig.~\ref{figure12} (master oscillator).
With that cavity configuration the threshold for the instability
onset is more easily reduced since FBG\ mirrors have no intrinsic losses.

Optimization of the FBG\ characteristics turns out to be crucial to obtain
both stable multimode operation and narrow spectral bandwidth
\cite{BordaisTh,Bordais03}. The selective low reflectivity FBG mainly
determines the smallest FWHM\ achievable in multimode operation ($\sim$1~GHz).
Most frequently, the range of launched pump powers where the double
FBG\ oscillators exhibit unstable behavior is very small. Sometimes no
self-pulsing is observed, even close to laser threshold. The FWHM of the
double FBG\ oscillator increases with the pump power. It can thus be adjusted
to optimize the OP\ performances of the laser. However, the pump power also
directly controls the power delivered by the oscillator. Therefore, to use it
as a master oscillator, care must be taken to get enough output power to feed
into the high gain power fiber amplifier. Recently, considerable progress has
been made in the processes used to imprint Bragg grating patterns onto
photosensitive fibers. Quality FBG tailored to the user's specifications are
now available. This, combined with a better control of all loss sources inside
the laser cavity (e.g., in the fiber splicings), allows for the routine
building of multimode tunable fiber oscillators with adjustable FWHM\ between
1 and 2~GHz.

Fig.~\ref{figure12} shows the complete structure of the dedicated all-fiber
MOPFA lasers designed for operation at 1083~nm with optimized spectral
characteristics in view of OP applications. The power booster is here, for
instance, a 5W fiber power amplifier with two-stage architecture. Linear
polarization is required at the output fiber connector. For the user's
convenience the output fiber may be a few meters long. Polarization
maintaining fibers are available but still quite expensive. A mechanical
polarization controller is thus included in the laser to pre-compensate for
polarization changes induced by light propagation in the output fiber.

Test measurements are performed in helium cells with a double FBG oscillator
in MOPFA configuration with a single-stage 2W fiber amplifier, and two other
ones with 5W fiber amplifiers. These lasers achieve broadband multimode
emission with, respectively, 1.63, 1.67 and 2.10~GHz FWHMs (measured at the
$\mathrm{He}$ resonance wavelength).

\begin{figure}[th]
%h=here, t=top, b=bottom, p=separate figure page
%
%
%
%
%
%
%
%
%
%
%
%
%
%
%
%
%
%
%
%
%
%
%
%
%
%
%
%
%
%
%
%
%
%
%
%
%
%
%
%
%
%
%
%
%
%
%
%
%
\par
\begin{center}
\leavevmode
\includegraphics[keepaspectratio,width=0.95\linewidth,clip= ]{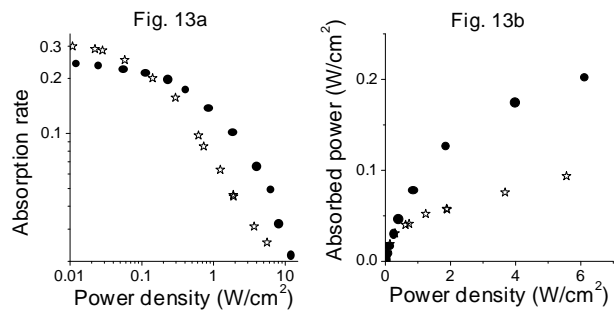}
\end{center}
\caption{Absorption rates (Fig. 13a) and deposited light power density (Fig.
13b) in the 1.6~mbar $^{3}\mathrm{He}$ cell, measured in identical conditions
for a 1.64~GHz FWHM double FBG oscillator (solid dots) and a DBR laser diode
(open stars) tuned to the C$_{9}$ line. }%
\label{figure13}%
\end{figure}

Fig.~\ref{figure13} displays absorption data for the 1.63~GHz FWHM fiber laser
with double FBG cavity. Comparison is made with concomitant measurements performed in
fully identical conditions with a singlemode laser diode. We use here the same
1.6~mbar $^{3}\mathrm{He}$ cell as that used to test the loop mirror
oscillator (Figs~\ref{figure10} and \ref{figure11}). However, for the present
series of measurements the discharge is stronger, leading to more metastable
atoms excited in the gas and proportionally higher absorption rates.
Fig.~\ref{figure13}a shows that at the limit of very low power densities (no
optical saturation) the absorption rate of the fiber laser is 0.77~$\pm$~0.02
times that of the DBR diode. This is in very good agreement with the expected
decrease obtained from the convolution of the laser and atomic line profiles
($1/\sqrt{1+L^{2}/D^{2}}$ = 0.772 for a laser FWHM $L~$= 1.63~GHz). Saturation
occurs at high powers, but clearly in a much less severe manner for the
multimode laser than for the singlemode one. The measured threshold power
density $\mathcal{P}_{\mathrm{1/3}}$ is much larger for the fiber laser
(3.1~W/cm$^{2}$) than for the DBR diode (0.6~W/cm$^{2}$). As a consequence the
absorbed light power can be as much as twice larger for the broadband fiber
laser at high power densities (Fig.~\ref{figure13}b).

\begin{figure}[th]
%h=here, t=top, b=bottom, p=separate figure page
%
%
%
%
%
%
%
%
%
%
%
%
%
%
%
%
%
%
%
%
%
%
%
%
%
%
%
%
%
%
%
%
%
%
%
%
%
%
%
%
%
%
%
%
%
%
%
%
%
\par
\begin{center}
\leavevmode
\includegraphics[keepaspectratio,width=0.95\linewidth,clip= ]{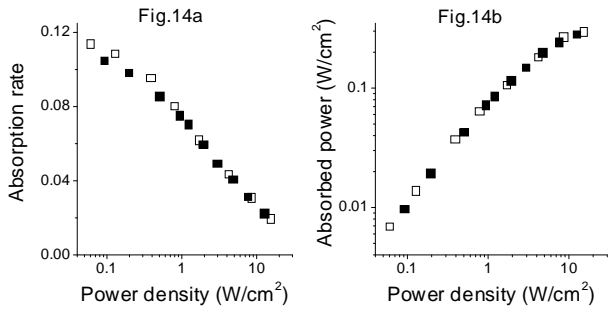}
\end{center}
\caption{Comparison of absorption measurements performed with 1.7 (solid
squares) and 2.1~GHz (open squares) FWHM multimode fiber lasers involving
double FBG cavity oscillators; 1.6~mbar $^{3}\mathrm{He}$ cell, C$_{9}$ light
with linear polarization.}%
\label{figure14}%
\end{figure}

Another set of comparative measurements is performed with the two double FBG
cavity\ oscillators in 5W MOPFA configuration that have 1.67 and 2.10~GHz
FWHMs (Fig.~\ref{figure14}). The absorption rate measured at very low power densities is 7\%
larger for the 1.67~GHz broad laser than for the 2.10~GHz one
(Fig.~\ref{figure14}a). This can be explained by the more favorable presence
there of a smaller bandwidth in the $L/D\approx1$ range due to the
$1/\sqrt{1+L^{2}/D^{2}}$ scaling. It is also in good agreement with both FWHMs
exceeding the best value for maximum absorption at low power densities as
suggested by the loop mirror oscillator tests and confirmed by the numerical
calculations ($L=$0.6-1~GHz, see Fig.~\ref{figure11}). No significant
difference between the two multimode lasers is observed above 1~W/cm$^{2}$.
This again matches the weak FWHM dependence of the test results and
expectations at high power densities (see Fig.~\ref{figure11}). Saturation
effects strongly increase above 1~W/cm$^{2}$ (Fig.~\ref{figure14}a), limiting
the rise of deposited power with incident laser power at higher photon flux
densities (Fig.~\ref{figure14}b).

The typical OP power densities presently used for production of polarized gas
lie between 0.2 and 2.4~W/cm$^{2}$
\cite{Becker94,Nacher99,Gentile01,LKBISMRM03,MainzISMRM03}. Our results
suggest that optimal performances should be achieved with 2~GHz-broad lasers
at the upper end of that power range, while 1.2-1.5~GHz FWHMs might be
preferred at the lower end. Preliminary OP measurements performed around
1~W/cm$^{2}$ indicate that the nuclear polarizations obtained in a 1.3~mbar
$^{3}\mathrm{He}$ cell at moderate discharge intensity are 10\% higher for the
1.67~GHz FWHM laser than for the 2.10~GHz FWHM one. Further experimental work
is still needed to reliably assess the optimal line width in multimode
operation with respect to the OP performances in well-controlled situations.
For this purpose, the double FBG cavity configuration provides the most
convenient opportunity to vary the laser FWHM ad libitum, from one to a few
GHz, through the control of the launched pump power. Systematic measurements
are in progress, both in sealed cells and in the presence of gas flow.
However, due the variety of OP conditions to be explored, efforts to develop a
predictive model based on accurate phenomenological input parameters must be
also continued to provide a comprehensive overview of the expected optimal OP performances.

\section{Conclusion}

We have demonstrated and tested the operation of all-fiber tunable lasers at
1083~nm for $^{3}\mathrm{He}$ OP applications. In the linear cavity
configuration, with a loop mirror and a low reflectivity fiber Bragg grating,
evolution of the laser line width and mode structure with launched pump power
has been observed and analyzed in detail using a variety of optical
measurements. In view of OP applications, simple tests have also been
presented to characterize the laser at the $\mathrm{He}$ wavelength on the
relevant time scales. The need for a multimode light source with a FWHM
matched to the atomic line width is clearly identified. It is met with the
double FBG cavity fiber oscillators.

These stable multimode oscillators with 1-2~GHz spectral FWHM are now combined
with power amplifiers in commercially available MOPFA devices, providing
stand-alone all-fiber laser sources delivering up to 10~W output power
\cite{Bordais03}. Compact turn-key 1083~nm lasers that require no servicing
from the user are attractive and convenient tools for research work. They are
also ideally suited for operation outside the physics laboratories, e.g., at
the clinics for onsite production of hyperpolarized gas next to the magnetic
resonance imaging scanner \cite{LKBISMRM03}. In a higher-stage MOPFA
configuration, a 1 or 2~W all-fiber MOPFA\ laser with double FBG\ oscillator
can be used as a master laser to launch a more powerful fiber power booster to
take full advantage of the truly multimode structure and allow very large
scale production of highly polarized $^{3}\mathrm{He}$ gas. Preliminary tests
of operation at 1083~nm up to 40~W have been performed \cite{WHprivatecom40W}.

For users, precise line width specifications and identification of the
ultimate physical limitations of MEOP will contribute to bringing new record
$^{3}\mathrm{He}$ nuclear polarizations within reach. For laser manufacturers,
the main technical challenge now lies in the cost-effective production of
robust, reliable and very powerful fiber sources for production of large
quantities of polarized gas to support the wide dissemination of established
applications of hyperpolarized $^{3}\mathrm{He}$ gas and the emergence of new
ones \cite{Medr1,Medr2,Meda,Phy1,Phy2,Phy3,Phy4,Phy5,Phy6,Phy7,Phy8}.

\end{document}